\documentclass[conference]{IEEEtran}
\IEEEoverridecommandlockouts
\def\BibTeX{{\rm B\kern-.05em{\sc i\kern-.025em b}\kern-.08em
    T\kern-.1667em\lower.7ex\hbox{E}\kern-.125emX}}
\usepackage{epsfig}
\usepackage{graphicx,xspace,xcolor,url}
\usepackage{times}
\usepackage{supertabular}
\usepackage{algorithmic,algorithm}
\usepackage{xspace}
\usepackage{multirow}
\usepackage{booktabs}
\usepackage{listings}
\usepackage{comment}
\usepackage{fancybox}
\usepackage{xspace}
\usepackage{paralist}
\usepackage{colortbl}
\usepackage{flushend}     
\usepackage{dblfloatfix}    
\usepackage{enumitem}
\usepackage{pbox}
\usepackage{array}
\newcolumntype{P}[1]{>{\centering\arraybackslash}p{#1}}

\usepackage{pifont}
\usepackage{url}

\usepackage[normalem]{ulem}
\newcommand{\eg}[0]{\textit{e.g.,}\xspace}
\newcommand{\ie}[0]{\textit{i.e.,}\xspace}
\newcommand{\etal}[0]{\textit{et al.}\xspace}
\newcommand{\xref}[1]{\S\ref{#1}}

\newcommand{\participantCount}{400\xspace}
\newcommand{\totalFunctions}{146\xspace}
\newcommand{\expertLow}{62\xspace}
\newcommand{\expertMid}{58\xspace}
\newcommand{\expertHigh}{26\xspace}
\newcommand{\informedLow}{20\xspace}
\newcommand{\informedMid}{57\xspace}
\newcommand{\informedHigh}{69\xspace}
\newcommand{\uninformedLow}{19\xspace}
\newcommand{\uninformedMid}{45\xspace}
\newcommand{\uninformedHigh}{82\xspace}

\newcommand{\uninformedParticipantCount}{200\xspace}

\newcommand{\uninformedResultperFunction}{100\xspace}

\newcommand{\uninformedCleanCount}{116\xspace}
\newcommand{\informedCleanCount}{94\xspace}
\newcommand{\totalDeviceTypes}{61\xspace}
\newcommand{\percentDecHighRisk}{60\%\xspace}

\newcommand{\sys}{Tyche\xspace}

\begin{document}


\title{\sys: A Risk-Based Permission Model \\for Smart Homes \vspace{-.5em}}

\author{
{\rm Amir Rahmati
}\\
Samsung Research America\\ Stony Brook University
\and
{\rm Earlence Fernandes}\\
University of Washington\\ University of Michigan
\and
{\rm Kevin Eykholt}\\
University of Michigan
\and
{\rm Atul Prakash}\\
University of Michigan
} 
 


\maketitle



\begin{abstract}
Emerging smart home platforms, which interface with a variety of physical devices and support third-party application development, currently use permission models inspired by smartphone operating systems---the permission to access operations are separated by the device which performs them instead of their functionality. Unfortunately, this leads to two issues: (1)~apps that do not require access to all of the granted device operations have overprivileged access to them, (2)~apps might pose a higher risk to users than needed because physical device operations are fundamentally risk-asymmetric---``door.unlock'' provides access to burglars, and ``door.lock'' can potentially lead to getting locked out. Overprivileged apps with access to mixed-risk operations only increase the potential for damage. We present \textit{\sys}, a secure development methodology that leverages the risk-asymmetry in physical device operations to limit the risk that apps pose to smart home users, without increasing the user's decision overhead. \sys introduces the notion of risk-based permissions for IoT systems. When using risk-based permissions, device operations are grouped into units of similar risk, and users grant apps access to devices at that risk-based granularity. Starting from a set of permissions derived from the popular Samsung SmartThings platform, we conduct a user study involving domain-experts and Mechanical Turk users to compute a relative ranking of risks associated with device operations. We find that user assessment of risk closely matches that of domain experts. Using this insight, we define risk-based groupings of device operations, and apply it to existing SmartThings apps. We show that existing apps can reduce access to high-risk operations by 60\% while remaining operable.


\end{abstract}


\section{Introduction}

Smart home platforms play the role of connecting heterogeneous devices and protocols while supporting third-party application development. Several such platforms are emerging, including Samsung SmartThings~\cite{stweb}, Apple HomeKit~\cite{homekitweb}, and Google Home~\cite{googleHome}. These platforms enable the promised home IoT benefits of better convenience, improved security, and more energy efficiency. However, such platforms pose security threats---Fernandes \etal performed an empirical evaluation of Samsung SmartThings, and discovered that its permission model automatically overprivileges apps~\cite{smartthings}. For instance, an app that locks doors after 9PM also gains the ability to unlock those doors at any time. Although overprivilege has traditionally plagued several types of computing models, in the context of IoT, it can cause physical and material harm, beyond the classically prevalent digital harm.


In this paper, using the popular and widely-used Samsung SmartThings as an example of a prototypical smart home platform, we explore the design space of permission models for smart homes with the goal of reducing the risk posed to users if their smart home apps are compromised or are malicious. In SmartThings, overprivilege occurs because its permission model groups device operations based on functional similarity---``door.lock'' and ``door.unlock'' both control the state of the lock. Functional grouping of operations helps reduce the cognitive burden on developers, when requesting permissions, and on users, when making security decisions during application installation. A very fine-grained model, where apps request permission for each device operation, only increases the burden on developers and users and has been shown to be impractical even in the smartphone environment~\cite{felt2012android}. The increased number of decisions a user has to make in a smart home environment makes this approach even less feasible. Therefore, despite the usability advantages of a functionally grouped permission model, attackers can use compromised or malicious apps, which are overprivileged, to cause security and privacy risks.  


We consider a middle ground between very fine-grained per-device-operation permissions and functionally grouped permissions. Our insight is based on the fundamental risk asymmetry of device operations---a property unique to physical devices. As an example, ``oven.on'' is a potential fire hazard, ``oven.off'' is potentially uncooked food, ``mic.on'' is a privacy risk, ``mic.off'' might only disable certain voice-assistant functionality like Amazon Alexa, ``door.unlock'' is a potential burglary risk, ``door.lock'' only locks the occupant outside (or inside) and is a potential annoyance. We present \sys, a secure development methodology that leverages this intuitive risk asymmetry, and groups physical device operations into equivalence classes of risk. In this risk-based model, apps specify access to device operations in terms of user-perceived risk-levels. Our goal is to retain the usability advantages of operation grouping while limiting the risks of overprivilege. 

\sys breaks down device functionalities into 3 risk categories (high, medium, and low). For our door lock example, ``door.unlock'' will be in a high-risk group, and ``door.lock'' will be in a medium-risk group for the door's operations. Developers request permissions by specifying the type of the device, and the risk category. So, an application that automatically locks the door after a specific hour will request for the permission ``door.medRisk'' and this gives it access to ``door.lock.'' If the application requests ``door.highRisk,'' it will gain access to ``door.lock'' and ``door.unlock.'' That is, there is an ordering relation between the risk categories. Intuitively, if the user is willing to give an application high-risk access to a device, the user is not concerned about low-risk access. The risk-based model is an improvement over functional grouping because it still groups operations, but in terms of risk. At the same time, \sys does not prohibitively increase the number of decisions user has to make, but ties an intuitive risk factor to them. Therefore, it bounds the damage that a compromised application can cause, and accurately communicates that potential to users at installation time. 

In designing \sys, we overcame several challenges: (1) We need a methodology to estimate the user-perceived risk of device operations. To that end, we introduce a survey instrument inspired by the methodology of Felt \etal that is designed to compute a relative ranking of user-perceived risk\cite{felt2012android}. The survey captures concrete risks that can arise from a variety of smart home devices, and then presents those risks to users. We use this methodology to obtain a ranking of user-perceived risk for Samsung SmartThings device operations. We focused on this platform because of its wide support for smart home devices (more than a 100). Furthermore, independently of the specific framework, physical device operations tend to remain the same. Therefore, our risk results are potentially applicable to other smart home platforms as well. (2) We need a technique to form equivalence classes of risk using the survey data. To that end, we utilize a clustering-based approach, and found it to work reasonably well in practice.




\noindent{\textbf{Contributions.}}
\begin{itemize}
\item We propose \sys, the first risk-based permission model for smart home platforms that leverages our insight that physical device operations are fundamentally risk-asymmetric.

\item We compute the user-perceived risk of device operations in the Samsung SmartThings platform using a user study of three domain experts and $\participantCount$ Mechanical Turk participants. This dataset is available at \emph{\url{https://github.com/Ethos-lab/datasets}}.

\item We empirically demonstrate that user-perceived risk of device operations closely matches that of the experts, motivating future research on permission prompt UI designs that focus on incorporating risk indicators.

\item Through a case study of three existing smart home apps, which we port to our risk-based permission model, we establish that the number of risky operations an attacker can issue reduces by $\percentDecHighRisk$.

\end{itemize}

We envision that \sys, including its user research methodology and permission model design, will serve as a set of guiding principles for developers of future smart home platforms.

\section{Background}
\label{sec:bg}
\subsection{SmartThings Framework}
\begin{figure*}[t!]
    \centering
    \includegraphics[width=\textwidth]{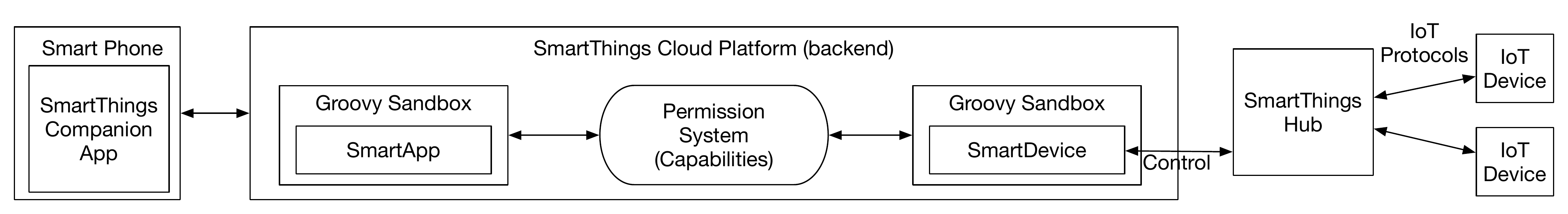}
    \caption{Samsung SmartThings Architecture.}
    \label{fig:smartthings}
    \vspace{-1em}
\end{figure*}

Figure~\ref{fig:smartthings} shows an overview of the SmartThings framework. It consists of: (1) a physical hub that users place in their home, (2) a proprietary cloud back-end that runs third-party applications, and (3) a companion application that runs on a user's smartphone that is used for local control and configuration. This smartphone app serves as a display device for the hub, as the hub itself has no display. Each hub supports a variety of home automation protocols such as ZWave and ZigBee. The companion application lets users download and install third-party SmartThings apps (or \textit{SmartApps}) into their SmartThings cloud account. The SmartApps, written in the Groovy programming language,\footnote{Groovy compiles to Java bytecode} interact with physical devices by issuing method calls (\eg lock a door) and by listening to events from the devices (\eg motion was detected). SmartApps are published to an application store. Physical devices are represented by \textit{device handlers} or \textit{SmartDevices}---pieces of code that use protocol-level instructions to communicate with devices. A device handler wraps a physical device, and exposes it to the rest of SmartThings. Although the platform provides device handlers for a wide variety of devices like door locks and speakers, developers can also write their own device handlers for unsupported devices.

\noindent{\textbf{SmartThings Permission Model.}} SmartApps must request capabilities (permissions)\footnote{Although SmartThings uses the word ``capabilities,'' they are in fact permissions and are not related to capability-based security. We will use ``permission'' in the rest of the paper.} for devices before they can interact with them. A capability is composed of a set of commands (method calls) and attributes (properties). Commands represent ways in which a device can be controlled or actuated. Attributes represent the state information of a device. For example, capability.lock contains the attribute ``lock,'' which represents the current state of the lock, and the commands: lock(), and unlock(), that an application can use to control the door lock. When a user installs a SmartApp, SmartThings triggers the device enumeration process where it lists all physical devices that support the requested capabilities. For instance, if a user's home has two motion sensors, and a SmartApp is requesting ``capability.motionSensor,'' then SmartThings will list both sensors in a permission granting UI. At that point, the user will decide whether to give the application access to the motion sensors or not.


\noindent SmartThings supports the functional grouping model---functionally similar device operations are grouped into a single capability. Third-party apps gain access at the level of such functionally-grouped capabilities. Depending on the app's functionality, this model can potentially lead to \textit{automatic} overprivilege~\cite{smartthings}.
Fernandes~\etal empirically evaluate the consequences of such overprivilege in their recent security analysis of SmartThings~\cite{smartthings}.
\subsection{Threat Model} 
We assume that SmartApps can be malicious or can be compromised. Fernandes \etal, in their recent security analysis of the SmartThings platform, demonstrated attacks of both types~\cite{smartthings}. The attacks depend on two factors: (a) users accept permission requests without realizing the risk an app poses; (b) apps can ask for more permissions than they need, or they automatically obtain more permissions than they need due to the design of the platform and the granularity of permissions. Under this assumption of overprivileged apps, our goal is to reduce the risk such apps pose to users. We consider other types of attacks to be outside the scope of this work. For example, a compromise of the platform itself will render permissions themselves useless. Orthogonal techniques are applicable to secure other parts of the smart home platform.


\section{\sys: The Risk-Based Permission Model}
\label{sec:design}
\vspace{-.3em}
Our key insight is that the risk asymmetry between functionally-related operations in smart homes creates an imbalance between the level of access to a device that an app needs, and the level of access provided to it by a traditional functionally-grouped access control system.
Permission grouping is vital from a usability standpoint. In theory, we could require that the application request access for each operation of a device. However, this extremely fine-grained system creates a poor user experience, and it ultimately results in users making poor security decisions due to fatigue~\cite{felt2012android}. We need to group operations together, but in the context of the smart home, this grouping can increase the risks users face.

\sys avoids this risk by assigning risk levels to each of a device's operations, and then groups operations of the device that have the same risk level into a group. An app can request access to a device's operations by specifying a risk level. The result is that we now have a way of communicating the risk of an app to the user, and we have a way of limiting the risk a compromised or malicious app poses to a user. 

There are three design challenges that we need to overcome to design a risk-based permission model:
First, how many risk levels should exist? In the simplest case, we could assign a risk level to each device operation. However, this can lead to the same decision fatigue that plagues extremely fine-grained access control systems. We believe that the answer to this question depends on the specific smart home platform, and the specific types of devices it supports. In the context of SmartThings, we experimented with three levels of risk (low, medium, high), and found it to be effective. \xref{sec:eval} contains an evaluation showing that with three levels of risk, existing apps can still function, but now with $60\%$ less access to high risk operations. Furthermore, as SmartThings supports more than a 100 types of devices covering a wide range, we envision that the results here are applicable to other platforms that work with similar devices.


Second, how do we create equivalence classes of device operations, grouped by risk level?  We envision such breakdown of risk to be done by smart home platform developers. To implement risk-based access control in current SmartThings platform, however, we asked three researchers with expertise in security of Internet of Things platforms to systematically evaluate every function currently available and put them in three risk categories. We averaged experts evaluations and use them as gold standard to group operations based on their risk. 

Third, how can we implement the risk levels on the SmartThings platform to concretely demonstrate their value? It is a closed-source system, and developers of SmartDevices (\ie device handlers) cannot define new capabilities. The straightforward way of implementing a risk-based system is to define new capabilities based on risk levels. However, this is not possible with SmartThings. To overcome this challenge, we adopt the approach used in ContexIoT~\cite{contexiot}. We propose automatically injecting security code into the apps at the source code level so that they communicate their risk level to users when they are installed. The remainder of the section details this process. It discusses our techniques by assuming a certain splitting of device operations. \xref{sec:eval} contains details on the exact splitting of SmartThings permissions into risk-based permissions. Our goal here is to discuss the methodology of enforcing risk-based permissions on SmartThings apps independently of a specific splitting of device operations. We stress that a risk-based permission model can be enforced in multiple ways. If a smart home platform were to allow customization (\eg analogous to Android OS modifications in smartphone research), then the platform itself could be modified to support risk-based permissions. 

\subsection{Developing a Risk-Based Permission Model}
As discussed in~\xref{sec:bg}, each SmartDevice in SmartThings can expose multiple capabilities (or permissions) depending on the kinds of operations the device supports. For example, a door lock will support operations related to battery management, lock control, and lock code programming. For each of these functional categories, there are associated permissions. \sys retains this model, but introduces a modifier that app developers need to use while requesting access to a device.

\noindent\textbf{Example.} Consider a battery management app that requests access to battery-backed devices around the home. Currently, such an app would make a permission request using the following line of Groovy code:\\

\noindent{$input$ $''battery\_device'',$ $''capability.battery''$}\\

If the user accepts the permission request, the app will gain access to all operations that are functionally grouped together inside capability.battery. As per our discussions, these operations may differ from a risk perspective. Under \sys, this app would instead specify a risk level as follows:\\

\noindent{$lowRiskRequest$\\
$input$ $''battery\_device'',$ $''capability.battery''$}\\

The first line is a method call that signifies that the next line of code is a permission request for low risk access to devices that support capability.battery. If a user accepts the eventual permission prompt, the app will gain access to all low risk operations defined by capability.battery. Thus, \sys keeps code changes to a minimum to help app developers move to a risk-based permission system. We envision that smart home platform developers could adopt a similar technique of using a single modifier for requesting risk-based access.

\subsection{Enforcing a Risk-Based Permissions Model}
\begin{figure*}[t!]    
    \centering
    \includegraphics[width=.8\textwidth]{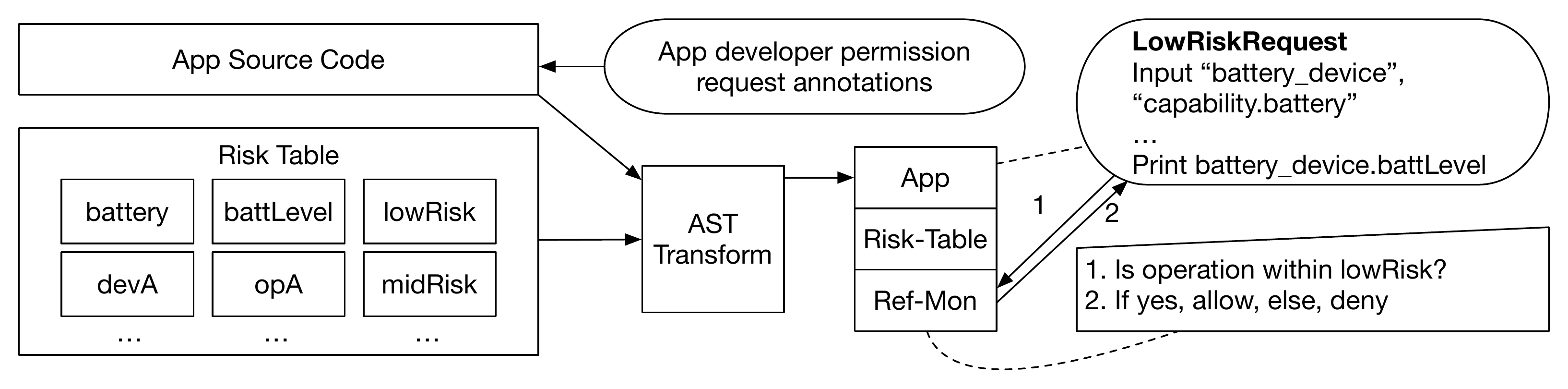}
    \caption{\sys rewriting architecture to enforce risk-based permissions.}
    \label{fig:rewriter}
    \vspace{-1em}
\end{figure*}

As the SmartThings platform is closed-source, \sys utilizes an app rewriting approach for enforcing our risk-based permission model. This is a common approach to enforcing security primitives in SmartThings. Notably, the platform itself performs rewriting to create a sandbox around apps (by allowing and denying certain operations)~\cite{smartthings}. Furthermore, recent work on enforcing contextual security policies on SmartThings uses the same approach~\cite{contexiot}. Rewriting occurs at the source-code level by operating on the Abstract Syntax Tree (AST) of the program. Groovy supports compiler extensions that perform rewriting of the AST. Our high-level approach is to: (1) Introduce a development-time annotation that the developer uses to specify the risk level of the access being requested; (2) a runtime reference monitor that the compiler injects into the app. This reference monitor hooks all device operation invocations, and verifies that the operation being invoked belongs to the risk level that was requested earlier. Figure~\ref{fig:rewriter} shows an architectural overview.

As input, the compiler extension takes app source code where the developer has annotated all permission requests with the risk rating. We envision this to be achievable with minimum developer effort as each permission is only requested once in each application. If there is a permission request without an annotation, the extension will throw a compilation error, guiding the developer to correctly annotate the permission request. The extension also takes in a risk table, that is simply a mapping of SmartThings capability names, operations (commands or attributes), and the risk rating. As we discuss in the next section, we use a user study to compute the risk rating. The extension then performs two steps: (1) it instruments all method invocations on objects that represent physical devices to first consult a reference monitor; (2) it injects a reference monitor into the app that is executed on every method invocation. Its job is to consult the risk table and determine if a particular invocation matches the previously requested risk rating. In our example, if the app attempted a high risk operation, the reference monitor would deny that because the developer had only requested low risk access.

To communicate the risk level of an app, our extension also injects startup code into the app. This results in a notification to the user of the risk levels that the app requested. The notification is shown on the corresponding smartphone app of SmartThings.

\begin{table}[tb]
\caption{Risk-Based Permission Examples. Commands are shown as $cmd()$ while attributes are presented as $attr$.}
\centering
\begin{tabular}{P{2cm}  P{1cm}  P{2cm}  P{2cm}  } \toprule

{\textbf{Permissions}} &
{\textbf{Low-Risk}}    &
{\textbf{Medium-Risk}} &
{\textbf{High-Risk}} \\ 
\midrule[\heavyrulewidth]
 
lock & -- & lock() & unlock(), lock \\ \midrule

alarm & -- & strobe(), siren(), alarm & -- \\ \midrule

switch & switch & -- & on(), off() \\ \bottomrule

\end{tabular}
\label{table:risk-based-perms}
\vspace{-2em}
\end{table}


\section{Usability Analysis}
\label{sec:eval}
We evaluate \sys from two aspects: First, to understand the usability of \sys, we evaluate user perception of risk in comparison with risk assigned by the domain experts by conducting a 2 stage user study using 210 participants. Second, we evaluate three applications under the \sys model. Our results show that \sys limits access to $60\%$ most high-risk operations, while neither decreasing the functionality, nor increasing user decision overhead. 

\subsection{User Study Setup}
To understand user perception of risk associated with physical device operations, we surveyed $\participantCount$ Mechanical Turkers.\footnote{We received exemption from our institution's IRB under title 45 CFR 46.101.(b). Survey data was collected anonymously and did not include any user-identifiable or sensitive information about the participants.} The survey started out by collecting basic demographic information. Per our IRB, we were not allowed to collect any user-identifiable data. The demographic information we collected are as follows: (1) Participant's age, (2) Number of people in their household, (3) The home platform they use, and (4) In cases where the participant's use SmartThings platform, the number of applications they have installed.

Next, the survey presented a description of the main task:

\begin{quote}
\textit{Imagine you have several apps installed in your home automation system. These programs require certain permissions and can perform actions on your behalf. Based on each action, indicate your level of concern based on the following Likert scale:}

\textit{
\begin{enumerate}
    \item \textbf{Not Concerned} - Select this for actions which pose little to no risk to you or the people living with you.
    \item \textbf{Mildly Concerned} - Select this for actions which pose a low risk to you or the people living with you, but are useful when used as intended.
    \item \textbf{Concerned} - Select this for actions which pose a medium risk to you or the people living with you, but are useful when used as intended.
    \item \textbf{Very Concerned} - Select this for actions which pose a high risk to you or the people living with you, but are useful when used as intended.
    \item \textbf{Would Not Allow} - Select this for actions which pose a high risk to you or the people living with you and are better left disabled.
\end{enumerate}
}
\end{quote}
We provided participants with a subset of device operations and asked them to rate their concern based on the above 5-point scale. We varied the information associated with the list of device operations based on the type of study condition. We had the following conditions:

\begin{table*}[tb]
\caption{Examples of the operations associated with a permission, their description, and their potential misuse scenario.}
\centering
\begin{tabular}{c c c m{5.5cm} } \toprule
\textbf{Permission}   & \textbf{Operation} & \textbf{Operational Description}                                       & \textbf{Misuse Scenario} \\ \midrule[\heavyrulewidth]
Switch & switch   & Track the status of the switch              & Publicly share when switch is turned on/off\\ \midrule
Switch & on()   & Turn on connected devices                                       & Turn on a device, such a space heater, for a long time, possibly causing a fire \\ \midrule
Lock & lock & Track the status of the lock & Publicly share when your door is unlocked \\ \midrule
Lock         & unlock() & Unlock door lock  & Automatically unlock doors without your knowledge     \\ \midrule
Lock         & lock() & Lock door lock  & Automatically lock doors without your knowledge     \\\bottomrule
\end{tabular}

\label{table:example}
\vspace{-1em}
\end{table*}


\begin{figure}
    \centering
    \includegraphics[width=.9\columnwidth]{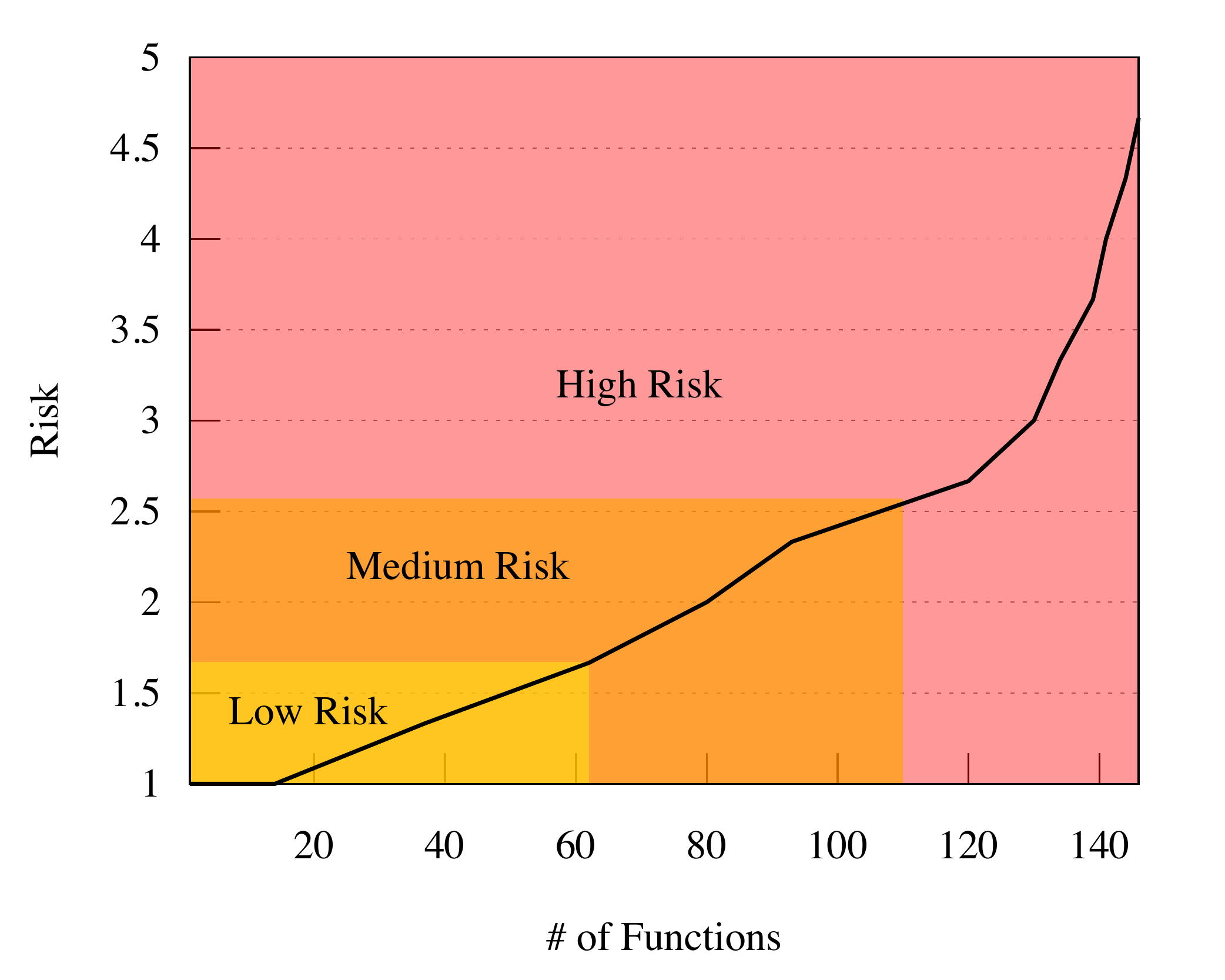}
    \vspace{-1em}
    \caption{Cumulative distribution of risk for the $\totalFunctions$ SmartThings functions assigned by domain experts. The distribution is divided to three ranges using K-mean clustering and is used as a basis for our risk-based permission model.}
    \label{fig:cdf}
\vspace{-1em}
\end{figure}

\begin{itemize}

\item \textbf{Uninformed User Ranking:}
We recruited $\uninformedParticipantCount$ users through Amazon Mechanical Turk to assess the risk of operations based on an \textit{operational description} only (Table~\ref{table:example}). To reduce decision fatigue, each user was provided with half of the device operations. In total, we obtained $\uninformedResultperFunction$ ratings per function from these users.

\item \textbf{Informed User Ranking:}
Similar to uninformed user study, we recruited $\uninformedParticipantCount$ users through Mechanical Turk. In addition to an \textit{operational description}, we provided each user with a \textit{misuse scenario} (Table~\ref{table:example}) to illustrate the potential risk associated with the device operation. Similar to the previous condition, we divided operations into two groups to reduce decision fatigue and collected $\uninformedResultperFunction$ ratings per device operation.
\end{itemize}

As a baseline for our study, We used the domain expert evaluation of all operations from Section~\ref{sec:design}. We consider this data as the gold standard for our system and compare other conditions to it.

\noindent \emph{Dealing with Random Clickers:}
Although prior studies have shown that results collected through crowdsourcing approaches are comparable to lab studies~\cite{kim2012filter}, our study is still susceptible to random clickers who answer the questions randomly to get the monetary benefit with minimum effort. To minimize this risk, we used our demographic questions as indicator questions to remove random responders (\eg family size = -2) and the Pearson's $\chi^2$ test to weed out random clickers as suggested by prior work~\cite{kim2012filter}.
These filters left us with total of $\uninformedCleanCount$ and $\informedCleanCount$ data points for our uninformed and informed condition, respectively.

\begin{table}[tb]
\caption{Age distribution of user study participants.}
\label{table:age}
\centering
\begin{tabular}{c c c c } \toprule
\textbf{Age}   & \textbf{$18-25$} & \textbf{$25-35$}                                       & \textbf{$35+$} \\ \midrule[\heavyrulewidth]
$\%$ of Participants & $24\%$ & $40\%$ & $26\%$  \\\bottomrule
\end{tabular}
\end{table}

\begin{table}[tb]
\vspace{-1em}
\caption{Number of Apps used by SmartThings users.}
\label{table:appcount}
\centering
\begin{tabular}{c c c c} \toprule
\textbf{\# of Apps}   & \textbf{$<5$} & \textbf{$5-10$}                                       & \textbf{$>10$} \\ \midrule[\heavyrulewidth]
$\%$ of SmartThings Users & $48\%$ & $43\%$ & $9\%$  \\\bottomrule
\end{tabular}
\end{table}


\subsection{User Study Results}
The age distribution of our participants is presented in table~\ref{table:age}. The population of our study consists of mostly young adults. We discuss how the age might affect user perception of risk in Section~\ref{sec:perception}. Table~\ref{table:appcount} also presents the number of apps SmartThings users install in their platform. Our results show that most smart home users stick to very few specific applications.
To assess the risk of each operation, each participant provides a Likert scale response (ranging from 1-5) as the risk of each operation. We average the score given by each group of our participants (uninformed, informed, and expert) and use K-means clustering (K=3) to divide these operations into three groups based on their risk. Table~\ref{table:stat} provides an overview of the K-means clustering result and the number of operations falling into the low-, medium-, and high-risk groups. When users consider the risk of an operation, they assign conservative risk values compared to domain experts. Informing the user about the actual misuse scenario lowers the perceived risk of some of the operations, but does not make the user complacent when compared to the expert. 

\noindent To examine the correlation between the risk assessment of experts (our gold standard) and users, we calculate the Pearson product-moment correlation coefficient (Pearson's $r$) between the experts and both uninformed and informed users results. Our results show moderate correlation ($0.6$) between experts and uninformed users and strong correlation ($0.75$) between the experts and informed users. These results motivate that a risk-based permission system can push users to make better risk assessments of access control requests from apps.

\begin{table}[!tb]
\vspace{-1em}
\centering
\caption{\# of operations that fall into each risk category \& the risk range based on the K-mean clustering. Making users consider the risk of an operation makes them more conservative than domain experts regarding safety of an operation.}
\begin{tabular}{c c c c}
      Risk         & Low    & Medium    & High   \\ \midrule[\heavyrulewidth]
\multirow{2}{*}{Uninformed User} & \uninformedLow       &  \uninformedMid      &   \uninformedHigh     \\
                & (1-1.91) & (1.92-2.42) & (2.43-5) \\ \midrule
\multirow{2}{*}{Informed User}   &  \informedLow      &   \informedMid     &   \informedHigh     \\ 
                & (1-1.94) & (1.95-2.59) & (2.6-5) \\      \midrule           
\multirow{2}{*}{Domain Expert}   &    \expertLow    &   \expertMid     &       \expertHigh \\
                & (1-1.83) & (1.84-2.83) & (2.84-5) \\ \midrule
\end{tabular}
\vspace{-2em}
\label{table:stat}
\end{table}




 



\vspace{-.5em}
\subsection{Case-Studies: Converting Apps to the Risk-Based Model}
\label{sec:casestudies}
We first compute the set of device operations accessible to three existing SmartApps based on the current SmartThings functional-grouping permission model, and then compare that set of accessible operations to a set we obtain after converting the apps to use our risk-based model. In general, we observe that \sys decreases applications access to risky operations by $60\%$ across all applications. Table~\ref{table:casestudies} shows the result of the conversion for our three case-studies.

\noindent{\textbf{LockDown}}
locks doors after a specified time of day. To perform its functionality, this application only needs access to ``lock()'' function. The current permission model, however, gives it access to the  ``unlock()'' function as well. If this application is compromised, an attacker can arbitrarily unlock the doors to the home. Under our risk-based model, the application only asks for low-risk access to the lock, that only grants the ``lock()'' operation to the app.

\noindent{\textbf{SmokeProtector}}
application sounds a siren if the smoke sensor is triggered. Under the current permission model, this application can access the ``off()'' operation. If compromised, an attacker can turn off the siren. Under our risk-based model, the application only asks for medium-risk access to the speaker. This results in only the ``siren()'' operation being accessible to the app.

\label{sec:EnergySaver}
\noindent{\textbf{EnergySaver}}
automatically turns off devices around the home if energy consumption is above a predefined value. Under the current model, this application can access the ``on()'' function of the switches. If compromised, it can turn on switches arbitrarily, thus powering any connected devices. This can negate the purpose of the application or, more dangerously, lead to a fire hazard by powering a device for an extended time, such as a space heater. Under our risk-based model, the ``off()'' and ``on()'' operations belong to the same high-risk group. Our users consider ``off()'' to be high-risk because it can disable safety-critical functions such as life support and refrigeration. Although our model does not separate the ``off()'' and ``on()'' operations, at least users are informed of the risks associated with using the appplication.
 We count the number of high-risk operations that apps can access in the current SmartThings permission model vs. the risk-based model. We find that on average, apps can remain functional and have access to $\percentDecHighRisk$ less high-risk operations. Therefore, if an application is compromised, our risk-based model reduces user risk compared to the current permission model.

\begin{table*}[tb]
\caption{We converted three existing SmartApps to our risk-based permission model. We find that only the EnergySaver application needs high-risk access to a switch. The other apps work without any high-risk access.}
\scriptsize
\centering
\begin{tabular}{P{2cm}  P{2.1cm}  P{2.3cm}  P{3.1cm}  P{2cm}} \toprule
 & \multicolumn{2}{c}{\textbf{Functional-Grouping Model}} & \multicolumn{2}{c}{\textbf{Risk-Based Model}} \\
 \cmidrule{2-5}
\textbf{App} & \textbf{Permissions} & \textbf{Accessible Operations} & \textbf{Permissions} & \textbf{Accessible Operations} \\ \toprule
Lockdown & \pbox{2.1cm}{lock,\\ contactSensor} & \pbox{2.3cm}{\textbf{attrs:} lock,\\ contact.\\ \textbf{cmds:} lock(),\\ unlock()} & \pbox{3.1cm}{lock.medRisk,\\ contactSensor.lowRisk} & \pbox{2cm}{\textbf{attrs:} contact.\\ \textbf{cmds:} lock()} \\ \midrule

SmokeProtector & \pbox{2.1cm}{alarm, \\ smokeDetector} & \pbox{2.3cm}{\textbf{attrs:} alarm,\\ smoke. \\
\textbf{cmds:} off(),\\ strobe(),\\ both(),\\ siren()} & \pbox{3.1cm}{alarm.medRisk,\\ smokeDetector.medRisk} & \pbox{2cm}{\textbf{attrs:} alarm, \\ smoke.\\ \textbf{cmds:} strobe(),\\ siren(), \\both() }  \\ \midrule

EnergySaver & \pbox{2.1cm}{powerMeter,\\ switch} & \pbox{2.3cm}{\textbf{attrs:} power,\\ switch.\\ \textbf{cmds:} on(),\\ off()} & \pbox{3.1cm}{powerMeter.medRisk,\\ switch.highRisk} & \pbox{2cm}{\textbf{attrs:} power,\\ switch. \\ \textbf{cmds:} on(), \\off()} \\

\bottomrule
\end{tabular}
\label{table:casestudies}
\vspace{-2em}
\end{table*}

\section{Discussion and Limitations}
\label{sec:discussion}

\subsection{\sys vs. ``Permission on first use''}
Smartphone platforms generally use either an application install-time permission request modality, or a first-use permission request modality. We surveyed four smart home platforms (SmartThings, AllJoyn, IoTvity, HomeKit), and observed that they currently use install-time permission requests. A potential reason for not including the first-use modality is that it assumes the presence of an interface to the user at all times. Smart home platforms in general might not have traditional user interfaces. For example, Samsung SmartThings only has a hub with no display. In this case, the platform uses an external graphical device, such as a smartphone, to surface the install-time permission request. However, such a device may not always be present at the time of a request. This works well for an install-time modality because the user installs apps with the help of a companion smartphone app. In the case of first-use permission requests, the smartphone may not be present when an application actually accesses a sensitive operation. \sys permissions can be used with either request modality, however, our prototype currently uses an install-time request.


\subsection{Perception of Risk}
\label{sec:perception}
A concern with using a risk-based access control system is that different people may have different perceptions of acceptable levels of risk. Although getting locked out may not be a major issue for an adult, it may be problematic for an elderly (or very young) person. Our user study for the prototype was conducted using adults only. Therefore, a limitation is that the resulting risk levels can potentially be biased towards adult perceptions of risk. An interesting future work direction is to examine risk perceptions of different age groups. However, \sys still provides benefits as it does not make applications more risky than they already are, and it still provides some notion of risk, even if the user's perceptions may not exactly match the system's risk levels.



\subsection{Overprivilege in \sys}
\sys does not completely eliminate overprivilege but grants applications a risk-based access privilege to each resource. This approach is intuitive---if a user trusts an application with high-risk functionality, they are fine allowing less sensitive functions as well. We observe that assigning risk levels to sensitive operations may or may not change the granularity of the associated permission's operations. This is in part due to our automated clustering process, and in part due to the need to avoid creating many risk levels (\eg A system with ten or fifteen different risk levels will most likely overwhelm the user and negate any benefits). For some apps (\eg EnergySaver  in~\xref{sec:EnergySaver}), this might not result in the app gaining even more limited access to sensitive operations---by a strict definition of overprivilege, even with a risk-based system like \sys, the app will still be overprivileged. Even in such cases, there are benefits to using \sys---it serves as a means to notify the user of the riskiness of a particular app.

\subsection{Surfacing Risk Levels to Users}
Our user study demonstrates the intuitiveness of a risk-based permission model. Yet, how best to present these risks to the user during installation time remains a challenge. Previous work has looked extensively at this problem in the context of smartphones~\cite{felt,felt2012android}. Our initial design in \sys used color coding and a second confirmation for high-risk permissions. We also limited the number of risk levels to three. We leave a user study on the effectiveness of various user interfaces and optimal number of levels as future work. An interesting challenge in surfacing permission requests to users in the context of smart homes concerns the interface mechanism. Hub-based platforms like SmartThings use a secondary device such as a smartphone for surfacing permission requests. Other platforms, such as voice-based assistants like Amazon Alexa or Google Home open up the possibility of a voice-based permission prompt and response mechanism. With a risk-based system, a challenge is to determine how best to leverage these newer modalities to communicate the risk level of an app to the user.

\section{Related Work}
Internet of Things platform security is an emerging research area. Fernandes~\etal~\cite{smartthings} performed a security analysis of SmartThings. They showed that more than $55\%$ of SmartThings apps are overprivileged and built four attacks that exploited the overprivilege. 
Various systems have since studied security issues in IoT platforms. FlowFence~\cite{flowfence,secdev} proposes data flow protection in IoT frameworks through the use of sandboxes and taint-tracking to impose flow control between data sources and sinks, and is concerned with ensuring that data is not misused. In contrast, risk-based permissions aim to limit the damage a malicious app can cause.
ContexIoT is a system that collects contextual information that it uses to help users make more informed decisions when granting permissions to apps~\cite{contexiot}. The contextual information is extracted using program analyses techniques. Such contextual prompting can further help in communicating \sys risk ratings to users. Similarly, Roesner \etal introduce Access Control Gadgets (ACGs) and its implementation on Android, LayerCake~\cite{layercake}, to improve contextual integrity. Our risk ratings can be made more contexual by adopting ideas from ACGs or from ContexIoT depending on the interface modality that is available to the smart home platform (\eg classic display, or voice-based interfaces). However, our goal in this work is to introduce a system design that is aware of risk asymmetries. Similar to ContexIoT, our work adopts a rewriting mechanism to enforce security decisions.


Current smart home platform permissions are modeled after smartphone permissions. There is a large body of work on analyzing and improving smartphone permissions~\cite{sokandroid}. For instance, Rahmati~\etal proposed using application context to guide access control decisions~\cite{contextAC}. While many of these approaches can be adapted for smart homes, they still overprivilege apps with high-risk operations because they do not take into account risk-asymmetry. 

Our user study methodology was inspired by Felt~\etal~\cite{felt}. While Felt's study focused on improving the warning messages in smartphone application installation UI, our study focuses on determining the risk level of operations, in order to define various access control levels in each device. Felt~\etal~\cite{felt2012android} also performed two user studies on the Android permission model. They found that that only $17\%$ of participants paid attention to permissions during installation and only $3\%$ could answer three permission comprehension questions correctly. These results motivate development of risk-based permission models, where users' understanding of risk factors closely matches domain experts. 

More generally, our work draws on results from risk-aware access control systems~\cite{riskrbac1,jaegerrisk,fuzzymls}. Adopting terminology from Petracca \etal, we are concerned with three type of risks: (1) risk due to authorizing unsafe operations;
(2) risk due to abuse of authorized permissions; (3) risk due to granularity of authorization hooks. Our insight is that current permission models in smart homes do not adequately communicate risk, resulting in users possibly authorizing unsafe operations simply because they do not know the risks involved. The second type of risk occurs due to the modern app model---apps can be malicious, or they can be compromised. The third type of risk occurs due to functional grouping of permissions. \sys tackles the first and third types of risk creating risk-similar groups of operations, and the communicating those risks to the user. We also introduce a human subjects methodology to estimate risk. However, \sys limits the effects of the second type of risk, instead of removing it altogether.

\sys currently uses static estimates of risk. However, risk can change depending on the user. 
We envision that results from fuzzy MLS systems could be applicable in encoding dynamic risk perceptions into a risk-based access control system~\cite{fuzzymls} for smart homes. We leave this to future work.

\section{Conclusion}
Smart home platforms currently use permission models inspired by smartphone OSes. These permission models group device operations based on functionality, and do not take risk-asymmetry into account. Although grouping operations together makes it easier for developers and users to work with permissions, it also overprilvileges apps. Due to the risk-asymmetry, overprivilege can drastically increase the potential for damage if apps are malicious or exploitable. Therefore, we introduced risk-based permissions as an alternative way to group device permissions. We designed \sys using app rewriting techniques to enforce risk-based permissions, and we conducted a study involving domain experts and Mechanical Turkers to compute user-perceived risk. Our study finds that experts and users share a similar perception of risk. Based on these findings, we re-grouped a physical device's operations into three groups---low-, medium-, and high-risk. We constructed such groupings for $\totalFunctions$ operations across $\totalDeviceTypes$ types of devices. We evaluated our risk-based model on three existing SmartApps. We found that these apps can be written in a way that reduces access to high-risk operations by $\percentDecHighRisk$.

 \noindent\textbf{Acknowledgements.}
We  thank  the  reviewers  for  their  insightful  feedback. This work was supported in part by NSF Grant No. 1646392 and 1740897, the MacArthur Foundation and the University of Washington Tech Policy Lab.


\bibliographystyle{acm}
\bibliography{references}

\end{document}